\documentclass[aps,prl,reprint,groupedaddress]{revtex4-1}

\usepackage{graphicx}
\usepackage{dcolumn}
\usepackage{amsmath}    
\usepackage{amssymb}
\usepackage{bm} 
\usepackage{hyperref}
\usepackage{latexsym}
\usepackage{verbatim}
\usepackage{color}
\usepackage{setspace}
\usepackage{ulem}
\usepackage{siunitx}
\def\beq{\begin{equation}}
\def\eeq{\end{equation}}

\def\beq{\begin{equation}}                           
\def\eeq{\end{equation}}                           
\def\bea{\begin{eqnarray}}                           
\def\eea{\end{eqnarray}}        

\begin{document}

\title{Speed inhomogeneity accelerates the information transfer in polar
flock}
\author{Sudipta Pattanayak$^{1}$}
\email{sudipta.pattanayak@bose.res.in}
\author{Jay Prakash Singh$^{2}$}
\email{jayps.rs.phy16@itbhu.ac.in}
\author{Manoranjan Kumar$^{1}$}
\email{manoranjan.kumar@boson.bose.res.in}
\author{Shradha Mishra$^{2}$}
\email{smishra.phy@itbhu.ac.in}
\affiliation{$^{1}$S. N. Bose National Centre for Basic Sciences, J D Block, Sector III, Salt Lake City, Kolkata 700106}
\affiliation{$^{2}$Department of Physics, Indian Institute of Technology (BHU), Varanasi, India 221005}

\date{\today}
\begin{abstract}

	{A collection of self-propelled particles (SPPs) shows coherent motion and exhibits
	 a true long range ordered (LRO) state in two dimensions. Various studies show that 
	 the presence of spatial inhomogeneities can destroy the usual long range ordering in the system. 
	 However, effects of inhomogeneity due to the intrinsic properties of the particles are barely addressed.
	 In this paper we consider a collection of polar SPPs moving with  inhomogeneous speed (IS) on 
	 a two dimensional substrate, which can arise due to varying physical strength 
	 of the individual particle. To our surprise the IS not only preserves the usual long range ordering present 
	 in the homogeneous speed models, but also induces  faster ordering in the system. Furthermore, The response of the flock 
	 to an external perturbation is also faster, compared to Vicsek like model systems, due to the frequent update of neighbors 
        of each SPP in the presence of the IS. Therefore, our study shows that the IS can help in faster information transfer 
	in the moving flock.}
\end{abstract}
\maketitle
\textit{Collective motion} of self-propelled particles (SPPs) or {\it flocking} 
is an ubiquitous phenomena in the nature where each constituent of the system 
shows a systematic motion at the cost of its internal energy \cite{Feder2007, Couzin2005, 
Couzin2013, Couzin2013PNAS, Read2016}. 
The flocks can vary in size from a few micrometers e.g., actin and
tubulin filaments, molecular motors \cite{Nedelec1997, Yokota1986}, unicellular organisms
such as amoebae and bacteria \cite{Bonner1998}, to several meters e.g.,    
bird flock \cite{Chen2019}, fish school \cite{Parrish1997} 
and human crowd \cite{Helbing2000} \textit{etc}. The Vicsek model \cite{Vicsek1995} 
is one of the most celebrated minimal model to understand the collective behavior of SPPs \cite{Vicsek1995}, 
and unlike its equilibrium counter part \cite{Mermin-Wagner}, the Vicsek model and other variants of this 
model exhibit the  existence of a true long-range ordered (LRO) state in two dimensions (2D) 
\cite{Vicsek1995, Chateprl2004, Chatepre2008, 
Toner1995, Toner1998}.\\
Majority of studies on these systems are performed with a clean substrate or in a homogeneous environment, 
however, natural systems are in general comprised of various kind of inhomogeneities 
\cite{ Morin2017, Chepizhko2013, Yllanes2017, Quint2015, Sandor2017, Reichhardt2017}. 
The recent studies show that the presence of spatial inhomogeneities breaks the true 
long range ordering in system \cite{Morin2017, Chepizhko2013, Yllanes2017, Quint2015, Sandor2017, Reichhardt2017, Rakesh2018, 
Toner2018E, Toner2018L}. The studies
of spatial inhomogeneities can help us to understand the escape dynamics and evacuation efficiency 
of crowd, e.g. the impact of obstacle on the efficient evacuation \cite{Lin2018, Dorso2011} 
or effective obstacle positioning near a narrow escaping door \cite{Zuriguel2016, ZuriguelJSM, Zuriguel2011}. 
All these studies are focused on 
the effects of spatially or external inhomogeneities, however, individual particles 
in a system may have different strength of energy extraction or physical strength which may act as intrinsic 
inhomogeneity. For example in a human crowd like the Kumbh Mela in India \cite{Kumbh} or the Hajj in Arabia \cite{Hajj}
and in a group of migrating animals each individual member has its self-propulsion speed depending on their 
physical strength. However, the effect of the inhomogeneous speed (IS)  amongst the SPPs   
has not been studied to the best of our knowledge, except in few recent experiments \cite{Lisicki2019, Goldstein2011} which 
show the existence of the inhomogeneity in speed of SPPs .\\
In this letter, we introduce a collection of SPPs where each of the SPP moves with a different self-propulsion speed or the IS. 
Surprisingly, we note that the presence of the IS accelerates the kinetics of ordering, and also
the ordered state is long-range in presence of the IS. More importantly, the response of the flock to an external 
perturbation is faster for larger IS  among the SPPs, since neighbors of each SPP are updated more frequently which 
leads to the faster information transfer inside the system. \\ 
\textit{Model:-} We consider a collection of $N$ polar SPPs moving 
on a two dimensional (2D)  substrate of size $L \times L$ with 
periodic boundary condition (PBC). 
Each particle is defined by its position 
${\bf r}_i(t)$ and orientation $\theta_i(t)$ and it moves with 
velocity $v_{0}^{i}{\bf n}_{i}(t)$, where ${\bf n}_{i}(t) = (\cos(\theta_i(t)), \sin(\theta_i(t))$ is the 
unit direction vector at time $t$ and $v_{0}^{i}$ is speed
of the $i^{th}$ SPP.
Unlike the previous studies \cite{Vicsek1995, Chateprl2004, Chatepre2008}, speed of the 
each particle is chosen from a Gaussian distribution and it remains 
fixed during the motion. SPPs try to follow their neighbors 
during the motion and interact among themselves through a 
short-range velocity alignment (ferromagnetic like) interaction
\cite{Vicsek1995}. 
We express the position and orientation update equations of the SPPs as:
\begin{equation}
	{\bf r}_{i}(t+1) = {\bf r}_{i}(t) + v_{0}^{i} {\bf n}_{i}(t),
\label{eqn_position}
\end{equation}
\begin{equation}
{\bf n}_{i}(t+1) = \dfrac{1}{W_{i}(t)}\bigg[\sum_{j \in R} {\bf n}_{j}(t) + N_{R}^{i}\eta_{0} \boldsymbol{\eta}_{i}\bigg],
\label{eqn_direction}
\end{equation}
The IS $v_{0}^i$ is obtained from a Gaussian distribution 
$P(v_0) = \frac{1}{\sigma \sqrt{2 \pi}}\exp[-\frac{1}{2}(\frac{v_{0}-\mu}{\sigma})^2]$, 
where $\mu$ is the mean and $\sigma$
is the standard deviation of the distribution. In our simulations,
we consider $\mu = 0.3$ and $\sigma$ is varied from $0.0$ to $0.05$. 
The direction of motion 
 of $i^{th}$ particle is calculated from the previous direction vectors of all  particles
inside its interaction range $R=1$.  $N_{R}^{i}$ is the number of neighbors within 
the interaction range $R$ of the $i^{th}$ particle. $\boldsymbol{\eta}$ is the random unit vector (noise) 
to incorporate the error made by the particle to follow its neighbors, and $\eta_{0}$ defines the 
strength of the noise. $W_{i}(t)$ is the normalization factor, which is norm 
of the vector inside the square bracket on R. H. S. of Eq. \ref{eqn_direction}. \\
 To investigate the {\it information transfer} and response to an external
perturbation, we introduce a small number of $N_{a}$ external agents 
in the steady state of the system. 
 External agents are immobile and placed randomly on the substrate with fixed  orientation 
 $\theta_{a}$. The SPPs interact with the  external agents  
 through the same short range alignment interaction defined 
in Eq. \ref{eqn_direction}. Due to the quenched orientation, these external
agents act like a small external field in the plane of the moving flock.  
The density of the external agents $\rho_a = \frac{N_{a}}{N}$ is one of the tunable
parameter.\\
We study the response of the flock to the external agents for various values of $\sigma$ of 
the IS distribution and density of external agents $\rho_a$.
The strength of the random noise $\eta_{0}$ is chosen to be $0.2$, such that the steady is 
an ordered state, and the density of SPPs $\rho_s=\frac{N}{L \times L}$ 
is kept fixed to $1.0$.  $10^5$ simulation steps are used to obtain the  steady state, and 
observables are averaged over $10-30$ independent realizations for better statistics. Each simulation 
step includes the updation of two update equations, Eq. \ref{eqn_position} and \ref{eqn_direction},
for all particles in the system. Total number of particles in the system is varied from $N=10^{4}$ to $10^5$.\\
\begin{figure}[t]
\centering
\includegraphics[width=1.0\linewidth]{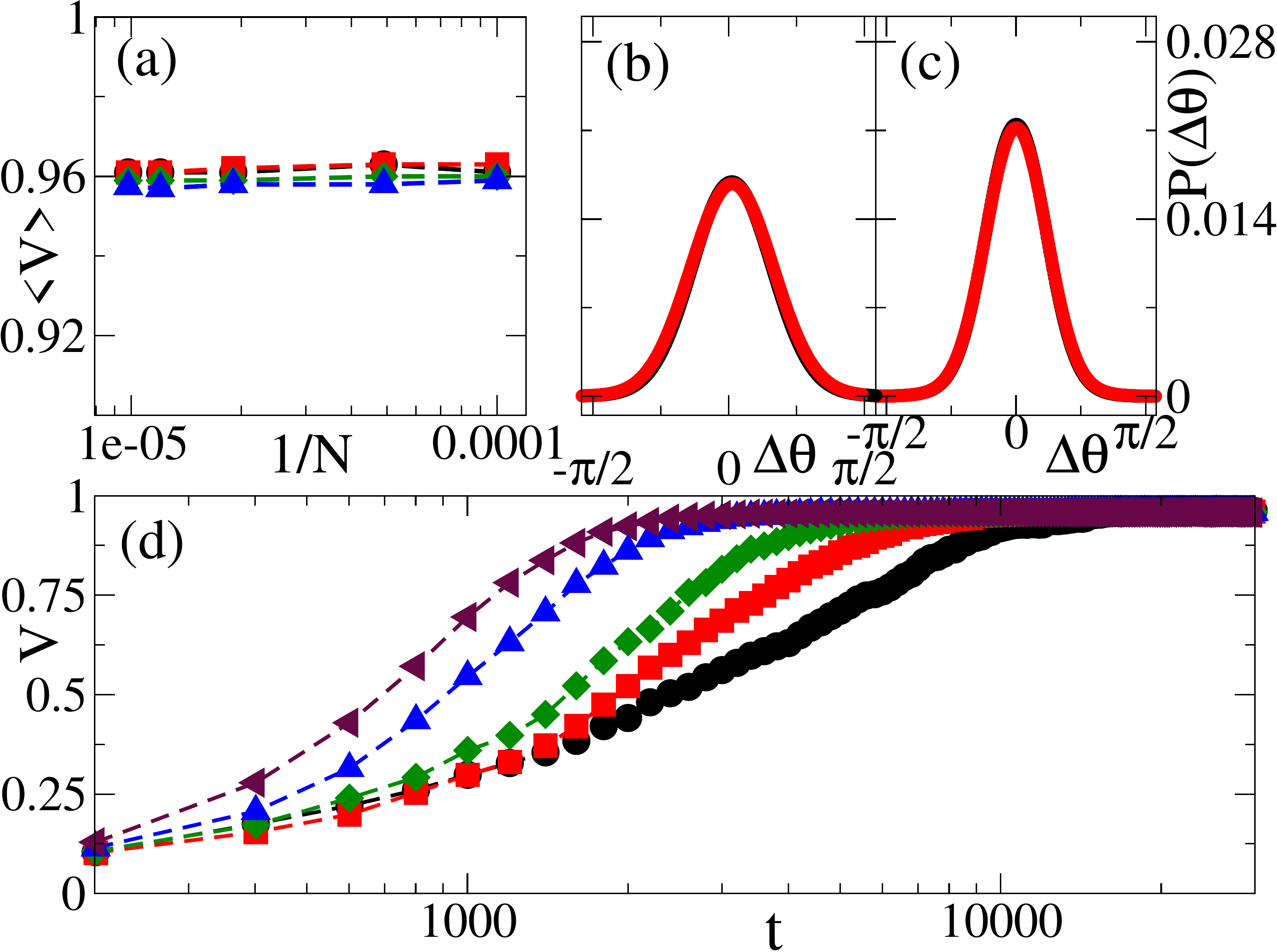}
	\caption{(Color online) (a) Plot of average global order parameter $\langle V \rangle$ \textit{vs.}
        $1/N$ is shown for different $\sigma$.
        Black circles, red squares, green diamonds,
        and blue triangles are for $\sigma = 0.0, 0.001, 0.01,$ and $0.05$, respectively.
	Error bars are less than symbol sizes. Plot of $P(\Delta \theta)$ of the SPPs at steady state
	for $\sigma = 0.0$ and $0.05$ are shown in (b) and (c), respectively. Black and 
	red circles are for $N=50000$ and $100000$, respectively.
	(d) The time series of $V$ is shown for different $\sigma$.
        Black circles, red squares, green diamonds, blue triangles up and maroon triangles down are
        for $\sigma = 0.0, 0.005, 0.01, 0.03$ and $0.05$ respectively. $N = 50000$.}
\label{fig_longrange_order}
\end{figure}
\textit{Ordered steady state:-}
In constant speed models or Vicsek like models \cite{Vicsek1995, Chateprl2004, Chatepre2008}
ordered state exhibits a true long-range order in 2D. In general the ordering in the system
 is measured by calculating the global order parameter which is defined as
\begin{equation}
        V(t)=\dfrac{1}{N}|\sum_{i=1}^{N}{\bf n}_{i}(t)|. 
\label{eqn_orderparameter}
\end{equation}
$V \sim 0$ for disordered state and it is close to 1 for completely ordered state.
We start the simulation with random position and orientation of the
SPPs, and with time, the system slowly evolves to the ordered state $V \sim 1$  for
$\eta_0=0.2$.
The $\langle V \rangle$ \textit{vs.} $1/N$ plot is shown in Fig. \ref{fig_longrange_order} (a),
 where $\langle .. \rangle$ denotes averaging over steady state time $5 \times 10^{4}$ to $10^5$ and $10$
independent realizations. We note that $\langle V \rangle$ remains independent of system size $N$, therefore,
one can
safely show the LRO state even for finite value of $\sigma$. 
Furthermore, we calculate
the probability distribution function (PDF) of the orientation of the particles $P(\Delta \theta)$ for different $\sigma$,
where $\Delta \theta$ is the deviation in the orientation of the particles from the
mean orientation direction. The width of $P(\Delta \theta)$  for
non-zero $\sigma$ does not change with the system size $N$, as shown in Fig. \ref{fig_longrange_order}(c).
It further confirms the LRO state of the system in presence of the IS similar to the constant speed models
 as shown in Fig. \ref{fig_longrange_order}(b).
The plot of $V(t)$ \textit{vs.} time $t$ is shown in Fig. \ref{fig_longrange_order}(d)
for four different values of $\sigma$ among the SPPs. The system takes less time to achieve the steady state
with increasing $\sigma$.
This behaviour of the system suggests that the IS
among the SPPs accelerates the ordering in the system. \\ 
\begin{figure}[t]
\centering
\includegraphics[width=0.9\linewidth]{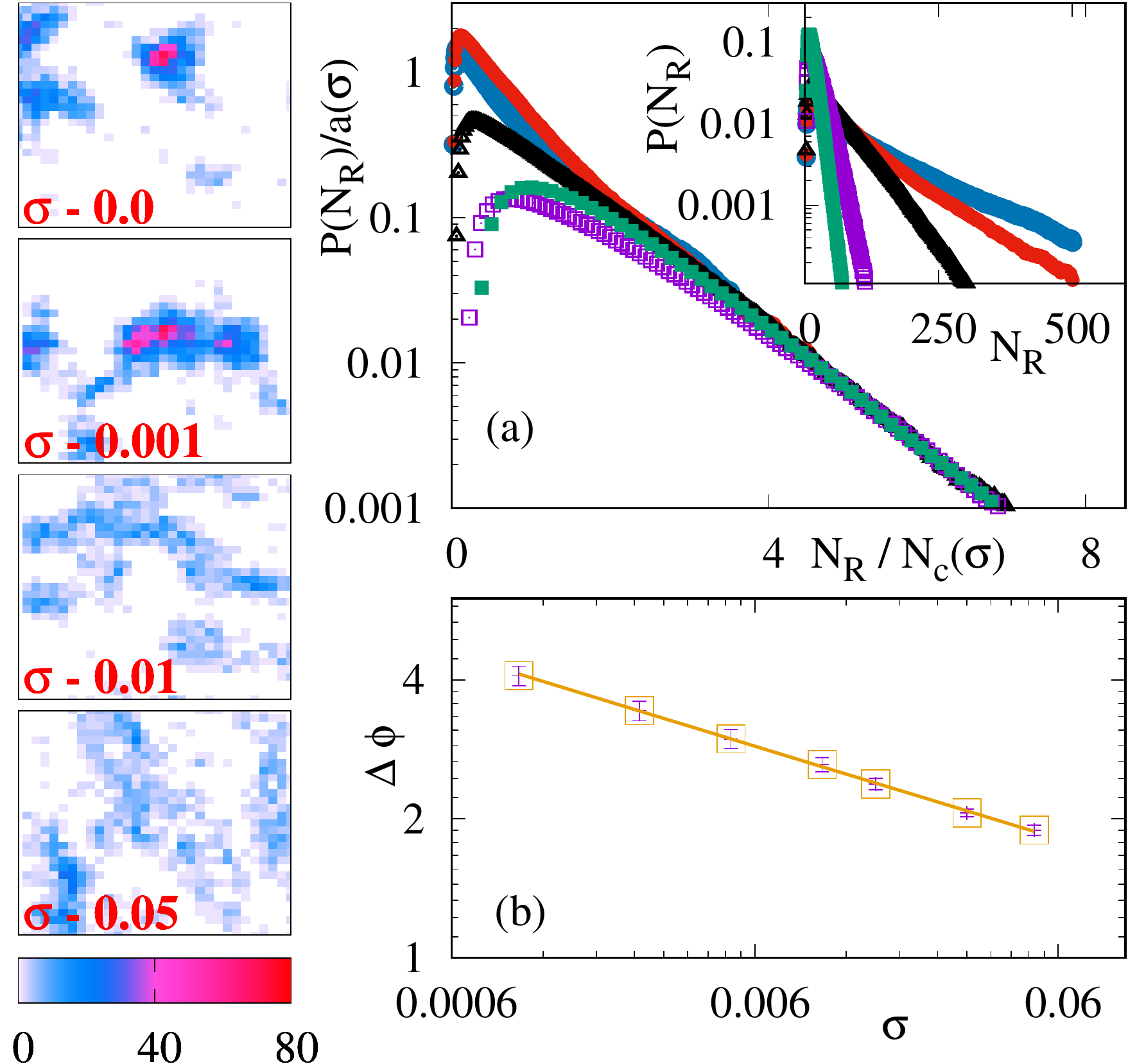}
	\caption{(Color online) \textit{Left column:} Snapshots of the system in the steady state is shown for different
        $\sigma$. $N = 20000$ and time $t=100000$. Color bar represents the number of particles in unit area.
	(a) Plot of $P(N_{R}/a(\sigma))$ \textit{vs.} $N_{R}/N_{c}(\sigma)$ 
	and $P(N_R)$ \textit{vs.} $N_R$ (\textit{inset}) for different $\sigma$ is shown. 
        Acme blue, red, black, violet, and teal color is for $\sigma = 0.0, 0.0005, 0.001, 0.01$ and $0.05$, respectively.
        $N = 100000$. (b) Variation of $\Delta \phi $ \textit{vs.} $\sigma$ is
        shown. Squares are numerical data, and solid line is the power law fitting
        of the numerical data with exponent $\sim -0.2$. Vertical lines shows the error bar.
        $N = 50000$.}
\label{fig_snapshots_initial}
\end{figure}
\begin{figure}[b]
\centering
\includegraphics[width=1.0\linewidth]{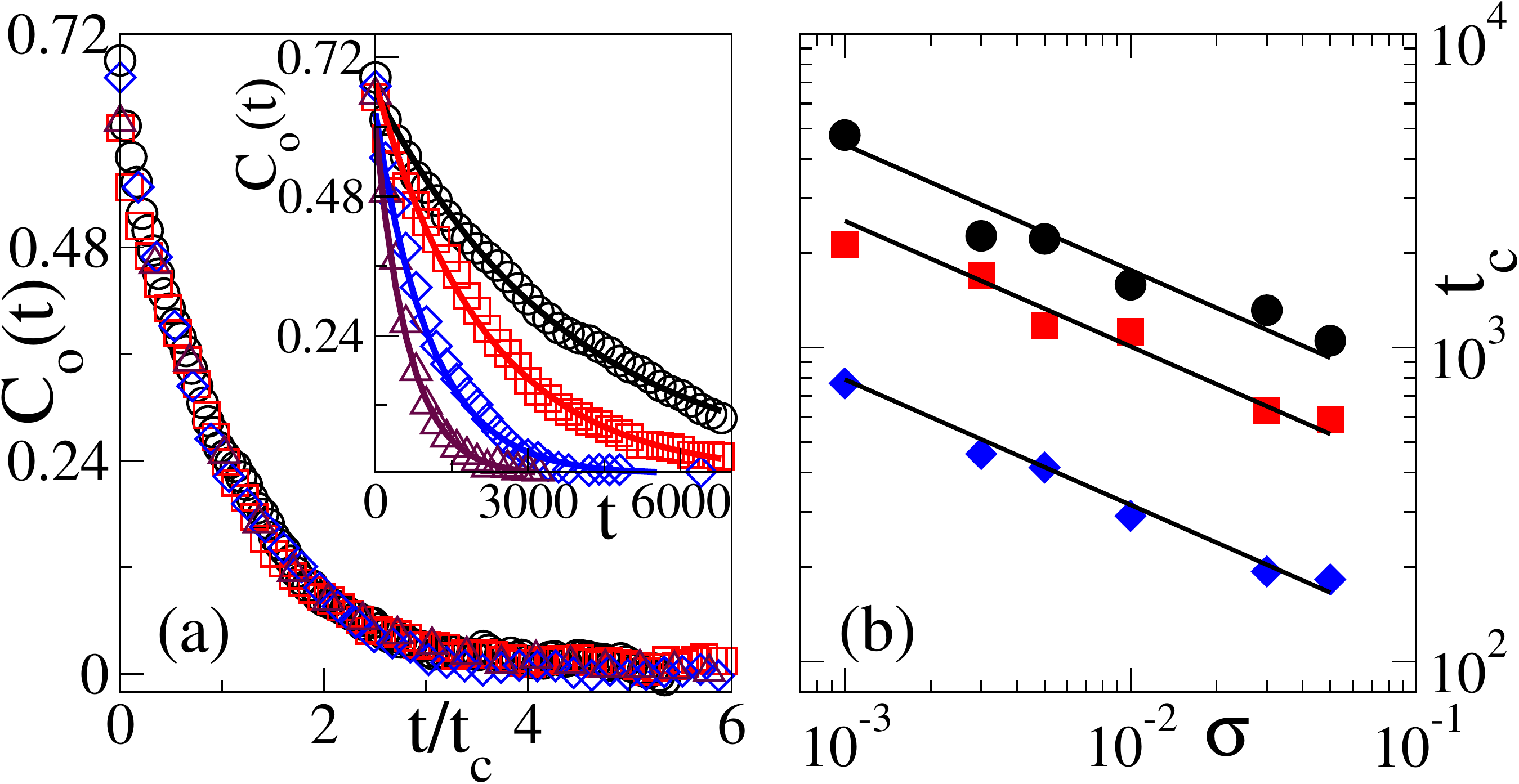}
        \caption{(Color online) (a) Plot of $C_{o}(t) $ \textit{vs.}
        $t/t_{c}$ is shown for different $\sigma$. \textit{Inset:}  Variation
        of $C_{o}(t)$ with time $t$ is shown for different $\sigma$.
        Black circles, red squares, blue diamonds, and maroon triangles up, are
        for $\sigma = 0.0, 0.001, 0.01$ and $0.05$, respectively. Solid
        lines in the inset are exponential fitting of the numerical data. Density
of external agents $\rho_a=0.005$.
        (b) The variation of $t_{c}$ 
        \textit{vs.} $\sigma$ is shown for three different
        values of $\rho_{a}$.
        Black circles, red squares, and blue diamonds are numerical data points
        for $\rho_{a}= 0.005, 0.01$ and $0.05$, respectively. Black solid line is
        power law fitting with exponent $-0.4$ of the numerical data points. Error bars
        are less than the symbol size.
        $N = 20000$.}
\label{fig_auto_corr}
\end{figure}
\begin{figure}[t]
\centering
\includegraphics[width=1.0\linewidth]{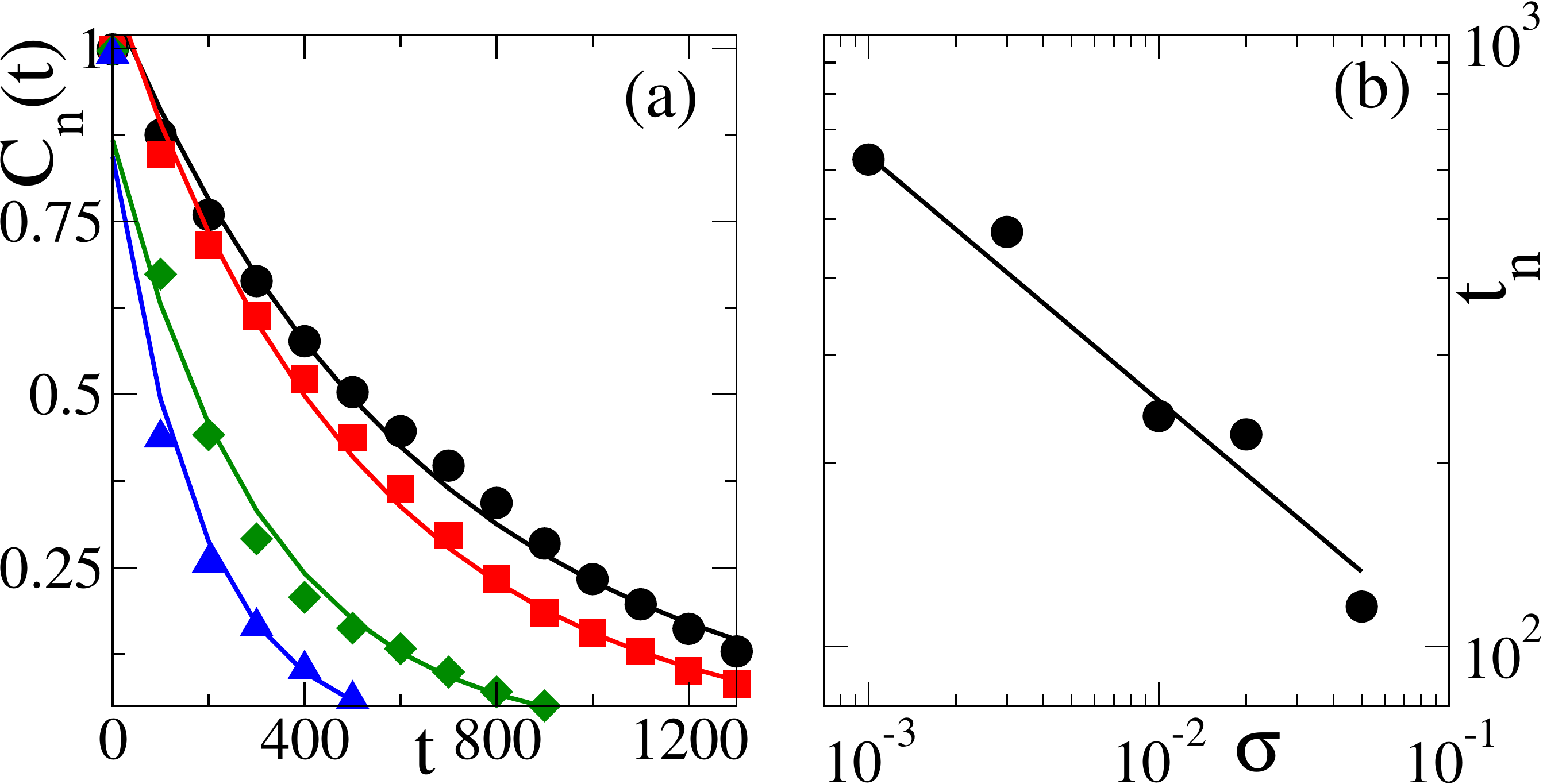}
	\caption{(Color online) (a) Plot $C_{n}(t)$ is shown for four different $\sigma$.
        Black circles, red squares, green diamonds, and blue triangles are for
        $\sigma = 0.0, 0.001, 0.01,$ and $0.05$, respectively. Solid lines are the exponential
        fitting of the
        numerical data.
        (b) Plot of $t_{n}$ \textit{vs.} $\sigma$ is shown.
         Black solid circles are numerical data and black solid line is
         the power law fitting of the numerical data with exponent $\sim -0.4$.
         Error bars are less than the symbol size. $N = 50000$.}
\label{fig6}
\end{figure}
\textit{Properties of the flock state:-}  We have already seen that 
the system with IS exhibits 
a LRO state. However, the steady state features of the ordered state
changes with increasing $\sigma$. In Fig. \ref{fig_snapshots_initial},
we show the real space snapshots of the system
for four different values of $\sigma(= 0.0, 0.001, 0.01$ and $0.05$) at time $t=10^{5}$. In the  constant speed model 
$\sigma=0.0$,  particles form isolated clusters which move coherently in 
one direction. But these isolated clusters break down and the system becomes  homogeneous with 
increasing $\sigma$. In the inset of Fig. \ref{fig_snapshots_initial}(a), we plot
the PDF of number of neighbors $P(N_{R})$ for different $\sigma$. $P(N_{R})$ \textit{vs.} $N_R$  
for $\sigma = 0.0$ decays with a long tail, but the tail sharpens with increasing $\sigma$. 
Hence, average number of neighbors for each SPP decreases with $\sigma$.  
 The plot of $P(N_R)/a(\sigma)$ \textit{vs.} $N_R/N_c(\sigma)$ is shown in Fig. \ref{fig_snapshots_initial}(a) (main), 
 where $N_c(\sigma)$ is obtained from 
the exponential fitting $a(\sigma) \times exp(-N_{R}/N_{c})$ of the tail of $P(N_{R})$ and 
$a(\sigma)$ is the pre-factor of the fitting function.
 To further confirm the effect of the IS on density clustering
 we calculate the  {\it density phase separation} order parameter $\Delta \phi$ 
 (the standard deviation of particles among the sub-cells). 
We calculate $\Delta \phi$ by dividing the whole $L \times L$ system into unit sized sub-cells.
$\Delta \phi$ of the system is defined as:
\begin{equation}
\Delta \phi  = \big \langle\sqrt{\dfrac{1}{L^{2}}\sum_{j=1}^{L^{2}}(\phi_{j})^{2}-(\dfrac{1}{L^{2}}\sum_{j=1}^{L^{2}}\phi_{j})^{2}} \big \rangle,
\label{eqn_density_phase_sep}
\end{equation}   
where $\phi_{j}$ is the number of particles in each sub-cell and 
$\langle .. \rangle$ represents averaging over $20$ realizations. 
We note that $\Delta \phi$ decreases with increasing $\sigma$ with a power $0.2$ as
shown in Fig. \ref{fig_snapshots_initial}(b). Hence, the 
system becomes homogeneous with increasing IS among the particles. \\
\textit{Inhomogeneous speed helps in faster information transfer:-} We claim that
the inhomogeneous speed distribution enhances the information transfer inside the flock, and
 a small number of quenched external agents are introduced in the system to characterize this behavior.
In the presence of the external perturbation the SPPs slowly reorient themselves along the
 direction of  the perturbation.
Response of the flock to  external agents is measured by direction auto-correlation function $C_{o}(t)$ of the SPPs: 
\begin{equation}
	C_{o}(t) = \langle cos(\theta_{i}(t)-\theta_{i}(0)) \rangle,
\label{eqn_auto_corr}
\end{equation}
where $\theta_{i}(t)$ and $\theta_{i}(0)$ are the directions of the $i^{th}$ particle at time
$t$ and at the reference time $t=0$ (when the external agents are introduced). 
$\langle .. \rangle$ denotes averaging over all the SPPs and $30$ different realizations.
$C_{o}(t)$ shows an exponential decay $e^{-t/t_c}$ with time, 
as shown in Fig. \ref{fig_auto_corr}(a)(inset). 
$t_c$ is the measure of time of a moving flock
to reorient its direction along the external perturbation. 
Furthermore, we also note that the auto-correlation function shows an excellent scaling behaviour 
for different $\sigma$, as shown in the main 
Fig. \ref{fig_auto_corr}(a).
The correlation time $t_{c}$ decays with a power $\sim 0.4$ with increasing $\sigma$, as shown in Fig. \ref{fig_auto_corr}(b). 
Therefore, the response of the flock to an external perturbation becomes faster
for IS distribution.
We also show that value of exponent remains invariant for different density of the 
external agents  $\rho_{a}= 0.005, 0.01$ and $0.05$ in Fig\ref{fig_auto_corr} (b).\\
Furthermore, we claim that the accelerated response is due to the {\it more frequent 
update} of neighbors for large IS. 
We calculate the change in neighbor list $Y(t)$ with time $t$ 
as defined in supplementary material (SM). 
We note that the 
neighbours list is updated more frequently for large IS. Furthermore, 
 we calculate the time lag $t$ neighbor auto-correlation function
\begin{equation}
	C_{n}(t)=\bigg\langle \dfrac{\sum_{t^{\prime}=1}^{T-t}(Y(t^{\prime})-\bar{Y})(Y(t^{\prime}+t)-\bar{Y})}{\sum_{t^{\prime}=1}^{T}(Y(t^{\prime})-\bar{Y})^{2}} \bigg\rangle,	
\label{nei_corr}	
\end{equation}	
where $\bar{Y}$ is the mean value of $Y(t)$ over the total time $T$ and $t<T$ \cite{Box1976}.
$\langle.. \rangle$ represents averaging  
over $15$ independent realizations. Faster decay of $C_n(t)$ means more frequent update of neighbour list.
We note that $ C_{n}(t) $ decays exponentially $e^{-t/t_n}$ with time $t$ 
as shown in Fig. \ref{fig6}(a). In Fig. \ref{fig6}(b), the algebraic decay   
of $t_{n}$ (with a power $\sim 0.4$) is shown as a function of $\sigma$, which  
is similar to the variation of $t_c$ with $\sigma$, as shown in Fig. \ref{fig_auto_corr}(b).  
The frequent update of neighbors of each SPP leads to the quicker 
information transfer.
To confirm that the two quantities, $t_{n}$ and $t_{c}$,
 are related, we show the movies
of the change in the neighbors list for a tagged particle and
the response of the SPPs to the external perturbation for $\sigma = 0.0, 0.001$
 and $0.01$ in \cite{sigma0}, \cite{sigma0.001} and \cite{sigma0.01},
 respectively. The movies show
that the SPPs slowly
reorient along the direction of the external perturbation. The
time required by the SPPs to orient along the external perturbation
decreases on increasing $\sigma$.
Also the neighbors of the particles change
more frequently for large $\sigma$, therefore,
 both $t_{c}$ and $t_{n}$ decrease with the IS. 
 \\
\textit{Conclusion:-}  We introduce a model for the collection of SPPs moving with the IS on a two-dimensional substrate, and  
	such inhomogeneous systems are abundant in the nature. However, effect of such inhomogeneity 
	is rarely studied in theory and simulations. In general the spatial inhomogeneity 
	into Vicsek like models destroy the long-range ordering \cite{Chepizhko2013, Morin2017}.  
        Surprisingly, our model with the IS preserves  
	the macroscopic LRO state found in the homogeneous or the constant 
        speed models\cite{Vicsek1995, Chateprl2004, Chatepre2008, Toner1995, Toner1998}. 
	Instead of destroying the LRO state, existence of the IS helps the system to reach the LRO state 
	faster compared to the constant speed models or the Vicsek like models \cite{Vicsek1995, Chateprl2004, Chatepre2008}. 
	Response of the flock to an external perturbation is measured by introducing a few orientationally and spatially
quenched external agents.\\
	We note that the IS of the SPPs enhances the response of the flock to the external perturbation. 
	The flock state becomes homogeneous with increasing the strength of the IS ($\sigma$), and it is 
	easier to bend the homogeneous cluster along the external perturbation. We show that
        the  $t_{c}$ and  $t_{n}$ decrease with increasing $\sigma$, 
        larger $\sigma$ enhances the update
	 in the neighbors and which in turn helps faster exchange of information inside the system. \\
       Our model may be useful to understand the effect of the IS on collective behavior in natural systems like 
	migrating birds or animals. Recently the 
experimental study by Lisicki \textit{et al.} on unicellular eukaryotes, e.g., flagellates
and ciliates, find that the probability distribution of swimming speed
of the eukaryotes does not follow the constant
speed model \cite{Lisicki2019}.
We hope this work will convince more scientists to consider intrinsic inhomogeneity
which helps in formation of flock state in active systems. \\
\textit{Acknowledgement:-}
We thank TUE computational facility at S.N.B.N.C.B.S.. S. Pattanayak thanks
Sriram Ramaswamy for useful discussions and also thanks Department of Physics IIT (BHU), 
Varanasi for kind hospitality during the visit. S. Mishra thanks DST, SERB(INDIA), project no. ECR/2017/000659 for partial financial support.


\begin{thebibliography}{9}
\bibitem{sigma0} Movie of a part of the system for $\sigma = 0.0$ is shown in
	the presence of external perturbation. Time $t = 0$ is the reference time, when the external agents are introduced in the steady
	state. The orientation of the external agents is $\theta_{a} = 0.087$ radians which is shown by a green arrow.
	The time span of the movie is $t=10^{4}$
	after the external perturbation is applied to the system at the steady state.
        Time difference between two consecutive snapshots
        is $t=20$.
	Cyan color represents the tagged particle. Circle represents
        the interaction radius of the tagged particle, and arrows indicate the direction of motion of the
        SPPs. Initially, the tagged
        particle is surrounded by the red particles and with time new neighbors appear (black particles).
	Also the direction of motion of the SPPs gradually changes along the external field ($\theta_{a} = 0.087$ radians). 
	$N = 40000$ and $\rho_{a} = 0.0125$.
\bibitem{sigma0.001} Movie of a part of the system for $\sigma = 0.001$ is shown in
        the presence of external perturbation. Time $t = 0$ is the reference time when the external agents are introduced in the steady
	state and orientation of the external agents is $\theta_{a} = 0.087$ radians which is shown by a green arrow. 
        The time span of the movie is $t=4.2 \times 10^{3}$
        after the external perturbation is applied to the system at the steady state. Time difference between two consecutive snapshots
        is $t=20$.
	Colors, circle and arrows represent same as in \cite{sigma0}.
        Systems parameters are same as in \cite{sigma0}.
\bibitem{sigma0.01} Movie of a part of the system for $\sigma = 0.01$ is shown in
        the presence of external perturbation. Time $t = 0$ is the reference time when the external agents are introduced in the steady
	state and orientation of the external agents is $\theta_{a} = 0.087$ radians which is shown by a green arrow.
        The time span of the movie is $t=8 \times 10^{2}$
        after the external perturbation is applied to the system at the steady state. Time difference between two consecutive snapshots
	is $t=10$.
        Colors, circle and arrows represent same as in \cite{sigma0}.
        Systems parameters are same as in \cite{sigma0}.
\end{thebibliography}
\end{document}